\documentclass{PoS}

\def\gsim{\mathrel{\rlap{\lower4pt\hbox{\hskip1pt$\sim$}}
    \raise1pt\hbox{$>$}}}         
\def\lsim{\mathrel{\rlap{\lower4pt\hbox{\hskip1pt$\sim$}}
    \raise1pt\hbox{$<$}}}         

\usepackage{multirow,cite}

\newcommand{\be}{\begin{equation}}
\newcommand{\ee}{\end{equation}}
\newcommand{\bea}{\begin{eqnarray}}
\newcommand{\eea}{\end{eqnarray}}
\newcommand{\bi}{\begin{itemize}}
\newcommand{\ei}{\end{itemize}}
\newcommand{\ben}{\begin{enumerate}}
\newcommand{\een}{\end{enumerate}}

\def\gev{\,\textrm{GeV}}
\def\tev{\,\textrm{TeV}}

\title{Parton Distributions at a 100 TeV Hadron Collider}

\ShortTitle{Parton Distributions at a 100 TeV Hadron Collider}

\author{\speaker{Juan Rojo}\\
        Rudolf Peierls Centre for Theoretical Physics, 1 Keble Road,\\
eUniversity of Oxford, OX1 3NP Oxford, United Kingdom\\
        E-mail: \email{juan.rojo@physics.ox.ac.uk}}

\abstract{
  The determination of the parton distribution functions (PDFs) of the proton will
  be an essential input for the physics program of a future 100 TeV
  hadron collider.
  The unprecedent center-of-mass energy will require knowledge
  of PDFs in currently unexplored
  kinematical regions such as the ultra
  low-$x$ region or the region of multi-TeV momentum transfers $Q^2$.
  In this contribution we briefly summarize the studies presented
  in the PDF section of the
  upcoming
  report on {\it ``Physics at a 100 TeV pp collider: Standard Model processes''}.
  First we map the PDF
  kinematical coverage in the $(x,Q^2)$ plane, quantify PDF uncertainties,
  and compute ratios of PDF luminosities between 100 TeV and 14 TeV.
  Then we show how
  the extreme kinematics of such collider lead
  to a number of remarkable PDF-related phenomena
  such as the top quark as a massless parton,
  an increased role of photon-initiated processes
  and the possible need of PDFs with high-energy resummation.

}

\FullConference{XXIV International Workshop on Deep-Inelastic Scattering and Related Subjects\\
         11-15 April, 2016\\
         DESY Hamburg, Germany}

\begin{document}

\paragraph{The structure of the proton at a 100 TeV collider.}
The accurate determination of the parton distribution functions (PDFs)
of the proton is
an essential ingredient of the LHC physics
program~\cite{Rojo:2015acz,Butterworth:2015oua,Accardi:2016ndt,
  Ball:2014uwa,Dulat:2015mca,Harland-Lang:2014zoa,Alekhin:2013nda},
and will be even more so at any future
higher-energy
hadron collider.
In particular, a new collider with center-of-mass energy of
$\sqrt{s}=100$ TeV, dubbed the Future Circular Collider (FCC),
would probe PDFs in currently unexplored kinematical regions,
such as the ultra low-$x$ region, $x\lsim 10^{-5}$, or the region of very large momentum
transfers, $Q^2 \ge (10~{\rm TeV})^2$.

 In this contribution we summarize the studies presented
  in the PDF section of the
  report on {\it ``Physics at a 100 TeV pp collider:
    Standard Model processes''}~\cite{report}.
  On the one hand, we quantify in various ways PDF uncertainties at a 100
  TeV hadron collider,  map their
  kinematical coverage in the $(x,Q^2)$ plane, and compute ratios
  of PDF luminosities between 100 and 14 TeV.
  On the other hand,
  we show how
  the extreme kinematics of such collider lead
  to a number of new phenomena
  related to PDFs such as the top quark as a massless parton,
  an increased role of photon-initiated processes
  and the need for PDFs with high-energy resummation.

\begin{figure}[t]
\centering
\includegraphics[width=0.93\textwidth]{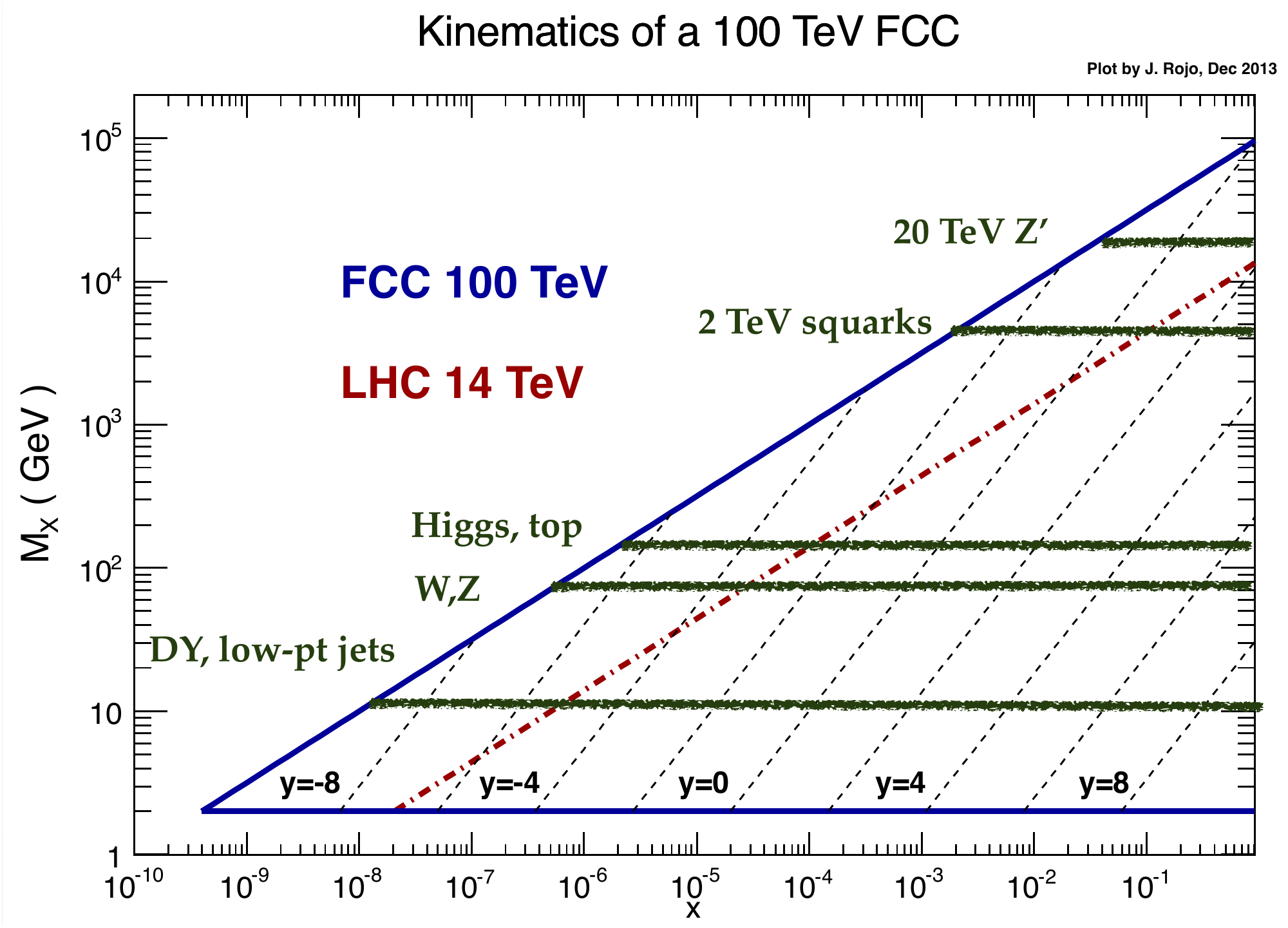}
\caption{\small Kinematical coverage in the $(x,M_X)$ plane of a
  $\sqrt{s}=100$ TeV 
  hadron collider (solid blue line),
  compared with the corresponding coverage of the LHC at
  $\sqrt{s}=14$ TeV (dot-dashed red line).
  The dotted lines indicate the lines of constant rapidity $y$ at the FCC.
  We also indicate the relevant $M_X$ regions for phenomenologically
  important processes, from low masses (Drell-Yan, low $p_T$ jets),
  electroweak scale processes (Higgs, $W,Z$, top), and possible new
  high-mass particles (squarks, $Z'$).
}
\label{fig:kinplot}
\end{figure}

In Fig.~\ref{fig:kinplot}
we represent the kinematical coverage in the $(x,M_X)$ plane,
where $M_X$ is the invariant mass of the produced final state,
for a $\sqrt{s}=100$ TeV
  hadron collider, 
  compared with the corresponding coverage of the LHC at
  $\sqrt{s}=14$ TeV.
  We also indicate the coverage  in  $M_X$ for phenomenologically
  important processes at the FCC, from low masses (such as Drell-Yan or low $p_T$ jets),
  electroweak scale processes (such as Higgs, $W,Z$, or top production),
  and hypothetical new
  high-mass particles (such as a 2 TeV squark or a 20 TeV $Z'$).
  
\begin{figure}[t]
\centering
\includegraphics[width=0.45\textwidth]{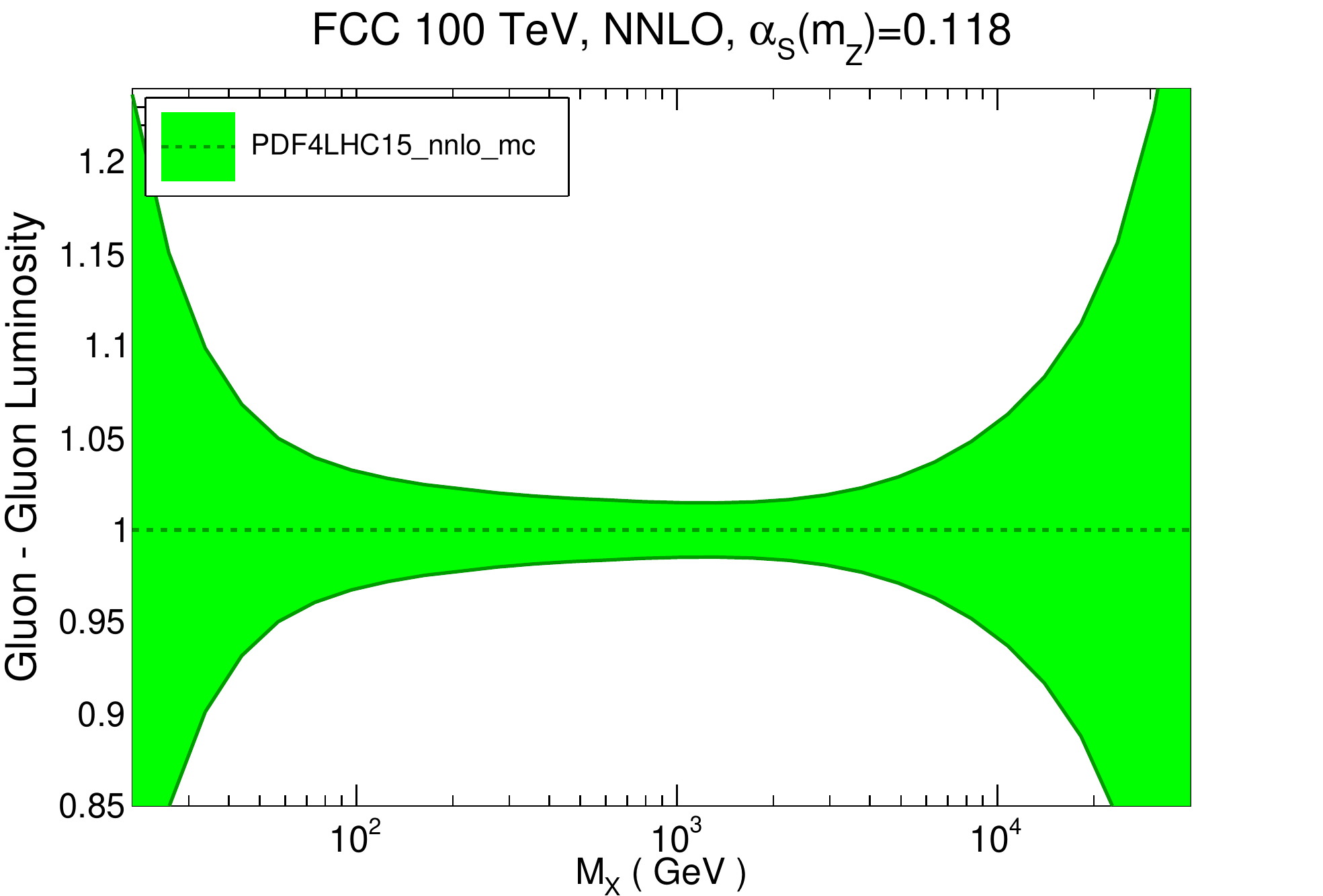}
\includegraphics[width=0.54\textwidth]{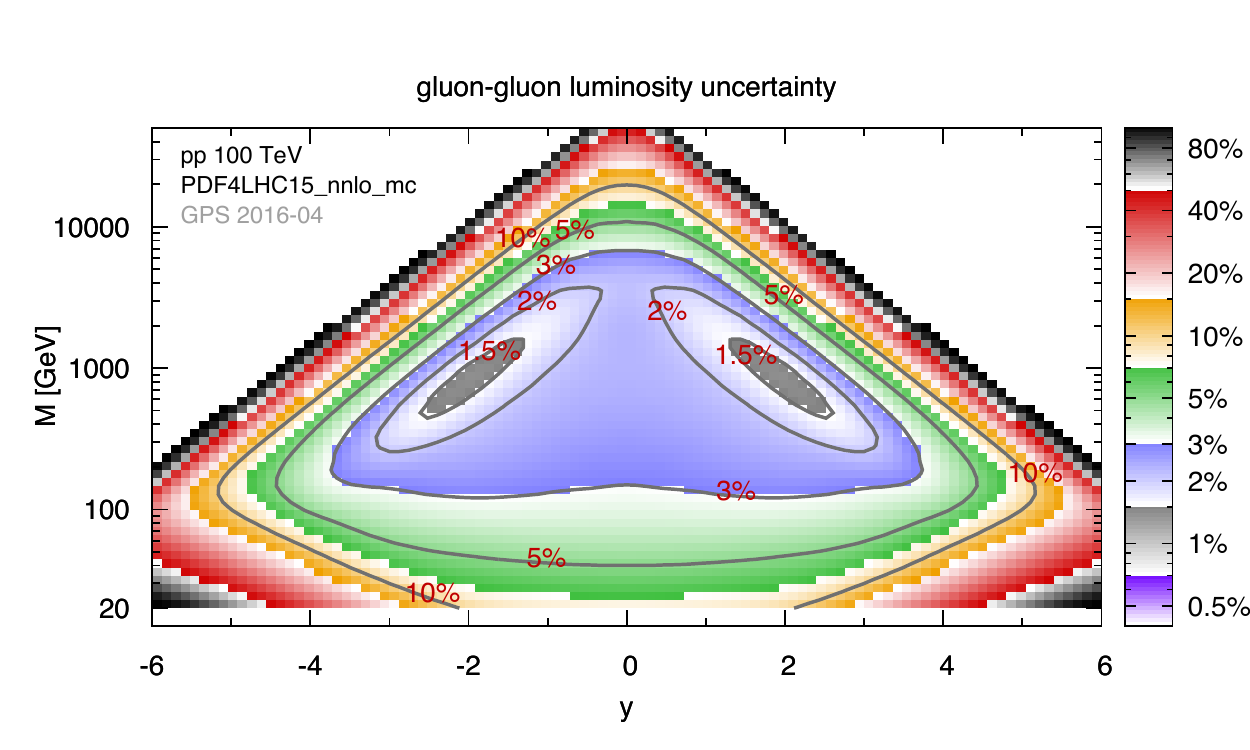}
\vspace{-0.7cm}
\caption{\small The PDF uncertainties in the gluon-gluon
  luminosity at the FCC with $\sqrt{s}=100$ TeV
  computed with the {\tt PDF4LHC15\_nnlo\_mc} set.
  Left plot: the rapidity-integrated luminosity
  $\mathcal{L}_{gg}(M_X)$.
  Right plot:
  the double-differential luminosity $\widetilde{\mathcal{L}}_{gg}(M_X,y)$.
}
\label{fig:lumicomp}
\end{figure}

  In Fig.~\ref{fig:lumicomp} we show the
  PDF uncertainties in the gluon-gluon
  luminosity at the FCC with $\sqrt{s}=100$ TeV
  computed with the {\tt PDF4LHC15\_nnlo\_mc}
  set~\cite{Butterworth:2015oua,Carrazza:2015hva}, both for
  the rapidity-integrated PDF luminosity
  ${\mathcal{L}}_{gg}(M_X)$ and for the corresponding
  double-differential
  luminosity $\widetilde{\mathcal{L}}_{gg}(M_X,y)$~\cite{report}.\footnote{We thank
G.~Salam for producing the double-differential luminosity plot.}
  PDF uncertainties are at the few-percent level for $100~{\rm GeV} \lsim M_X \lsim 5$ TeV,
  increasing for larger values of $M_X$, relevant for heavy particle
searches, and for smaller values of $x$, relevant for electroweak physics
and semi-hard QCD.
The same qualitative behaviour is observed for other
initial-state partonic combinations.

\begin{figure}[t]
\centering
\includegraphics[width=0.60\textwidth]{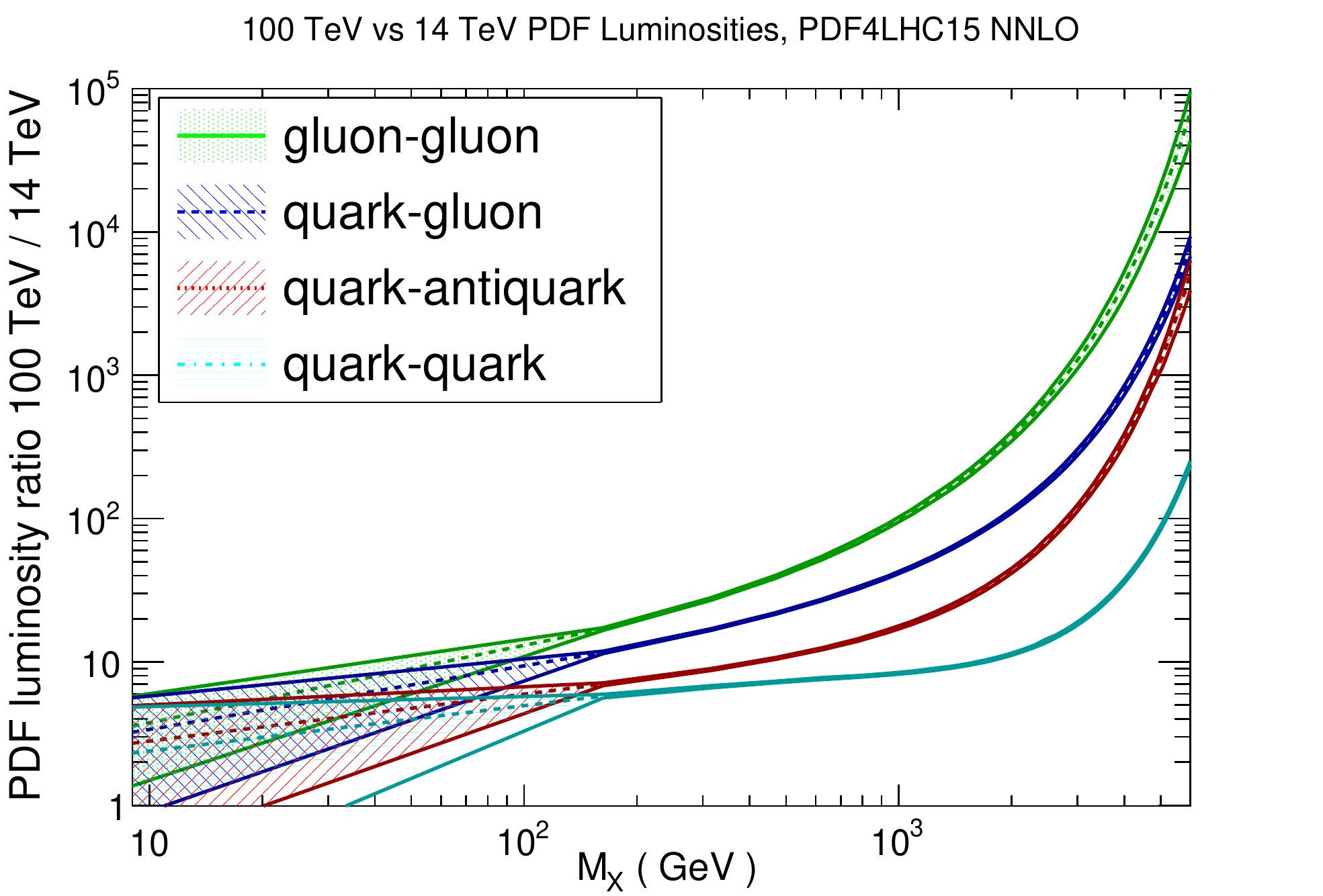}
\caption{\small The ratio of PDF luminosities
   between $\sqrt{s_1}=100$ TeV
   and  $\sqrt{s_2}=14$ TeV for different initial-state channels,
   computed with the PDF4LHC15 NNLO set.
\label{fig:lumirat} }
\end{figure}

Next we compute the ratio of the
rapidity-integrated PDF luminosities between 100 TeV and 14 TeV, for
different initial-state partonic channels,
$\mathcal{L}^{(100)}_{ij}(M_X)/\mathcal{L}^{(14)}_{ij}(M_X)$.
These ratios provide a direct method to rescale production cross-sections
between 14 and 100 TeV for processes dominated by a single
initial-state luminosity.
In Fig.~\ref{fig:lumirat} we show the ratio of PDF luminosities
between $100$ TeV
  and $14$ TeV for  different initial-state channels,
  using as input {\tt PDF4LHC15\_nnlo\_mc}, with the corresponding
  68\% CL PDF uncertainties.
We observe that for low invariant masses, $M_x \lsim 100$ GeV,
the increase in parton luminosities when going from the LHC to the FCC is
moderate, a factor 10 at most.
On the other hand, the luminosity ratio increases rapidly as we move away
from the electroweak scale, since these the increase in energy of the FCC
dramatically dominates over the large-$x$ fall-off of the PDFs at the LHC.
For invariant masses around $M_X\simeq 1$ TeV, for instance, the $gg$, $qg$,
$q\bar{q}$ and $qq$ luminosity ratios are $\simeq 100, 50, 20$ and 10,
respectively.
In general, gluon-initiated processes are those that will benefit more from the increase
in center-of-mass energy due to the rapid rise of the gluon
PDF at medium- and small-$x$ from DGLAP evolution.

\paragraph{Photon-initiated processes at 100 TeV.}
Consistent evaluation of electroweak
corrections require PDFs with QED effects including
a determination
of the photon PDF $\gamma(x,Q)$.
A number of QED PDF sets are available: MRST2004QED~\cite{Martin:2004dh}, 
NNPDF2.3QED~\cite{Ball:2013hta} and the recent
CT14QED~\cite{Schmidt:2015zda}, and PDF evolution
with QED effects has been implemented in the
{\tt APFEL} public PDF evolution
program~\cite{Bertone:2013vaa}.
Remarkable, already at the LHC, photon-initiated diagrams can be comparable
to quark-initiated ones, and this trend continues at the FCC energies,
and to illustrate this now
 we provide predictions for electroweak production
processes at $\sqrt{s}=100$ TeV.
The results have been obtained with
{\tt aMC@NLO}~\cite{Alwall:2014hca} using the \texttt{apfel\_nn23qednlo0118\_lept}
PDF set~\cite{Bertone:2015lqa}.

\begin{figure}[t]
  \centering
  \includegraphics[clip=true, trim=0.cm 3.5cm 0.7cm 1cm, width=0.49\textwidth]{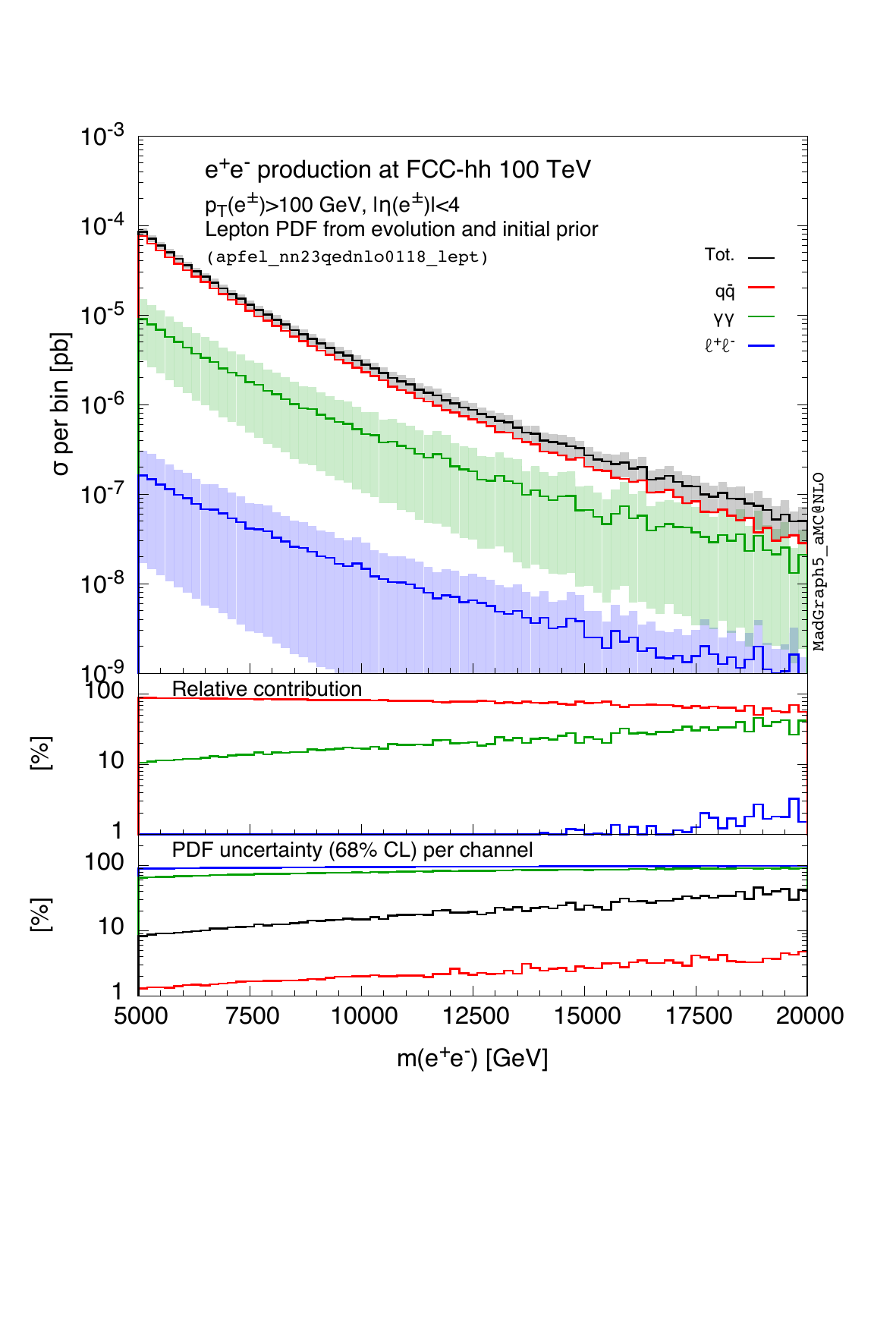}
   \includegraphics[clip=true, trim=0.cm 3.5cm 0.7cm 1cm, width=0.49\textwidth]{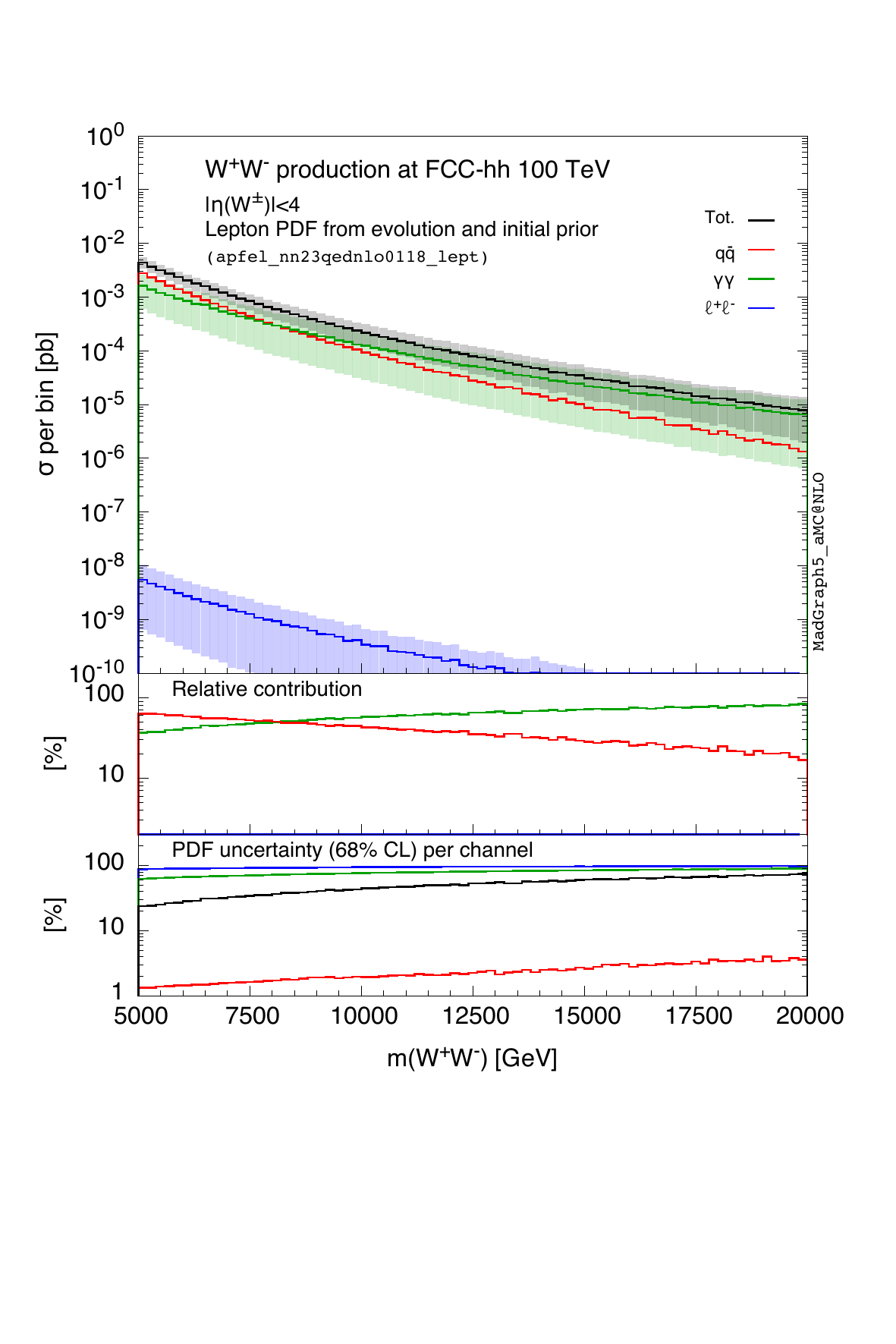}
  \caption{\label{fig:epem-fcc1}
    The invariant mass distribution of dileptons in $e^+e^-$ (Drell-Yan) production
    (left) and for $W^+W^-$ pairs (right) at a  100 TeV hadron collider, separating
    the contribution from each initial state.
  }
\end{figure}

In Fig.~\ref{fig:epem-fcc1} we show the invariant mass distribution of
lepton pairs in neutral-current Drell-Yan production at $\sqrt{s}=100$ TeV
for  $m_{e^+e^-} \ge 5\tev$, separated
in the different initial-state contributions, where the leptons must satisfy acceptance
cuts of  $p_T^{e^\pm} \ge 100~\gev$ and $|\eta_{e^\pm}| \le 4$.\footnote{We thank V.~
  Bertone, S.~Carrazza,
D.~Pagani and M.~Zaro for providing this plot.}
    The lower panel shows the corresponding
    PDF uncertainties.
   We find that  the photon-initiated contribution
is $\ge 10\%$ for all the invariant mass range, although
with large associated uncertainties.
One of the reasons for this is that in the DY process the
$q\bar{q}$-channel receives an additional kinematic suppression due to
$s$-channel diagrams that are absent in the
$\gamma\gamma$-channel.
In Fig.~\ref{fig:epem-fcc1} we also show the
differential distributions for the
invariant mass of the di-boson pair $m_{W^+ W^-}$
in $W^+W^-$ production.
We find that the
photon-initiated contribution could dominate over
the quark-antiquark annihilation for  $m_{W^+ W^-} \ge 7.5$ TeV,
also here with substantial PDF uncertainties.
The results of Fig.~\ref{fig:epem-fcc1} highlight the importance of
photon-initiated processes at high-energy hadron colliders,
and the need to reduce the theoretical
uncertainties that currently affect determinations
of the photon PDF.

\begin{figure}[t]
  \centering
  \includegraphics[width=0.47\textwidth]{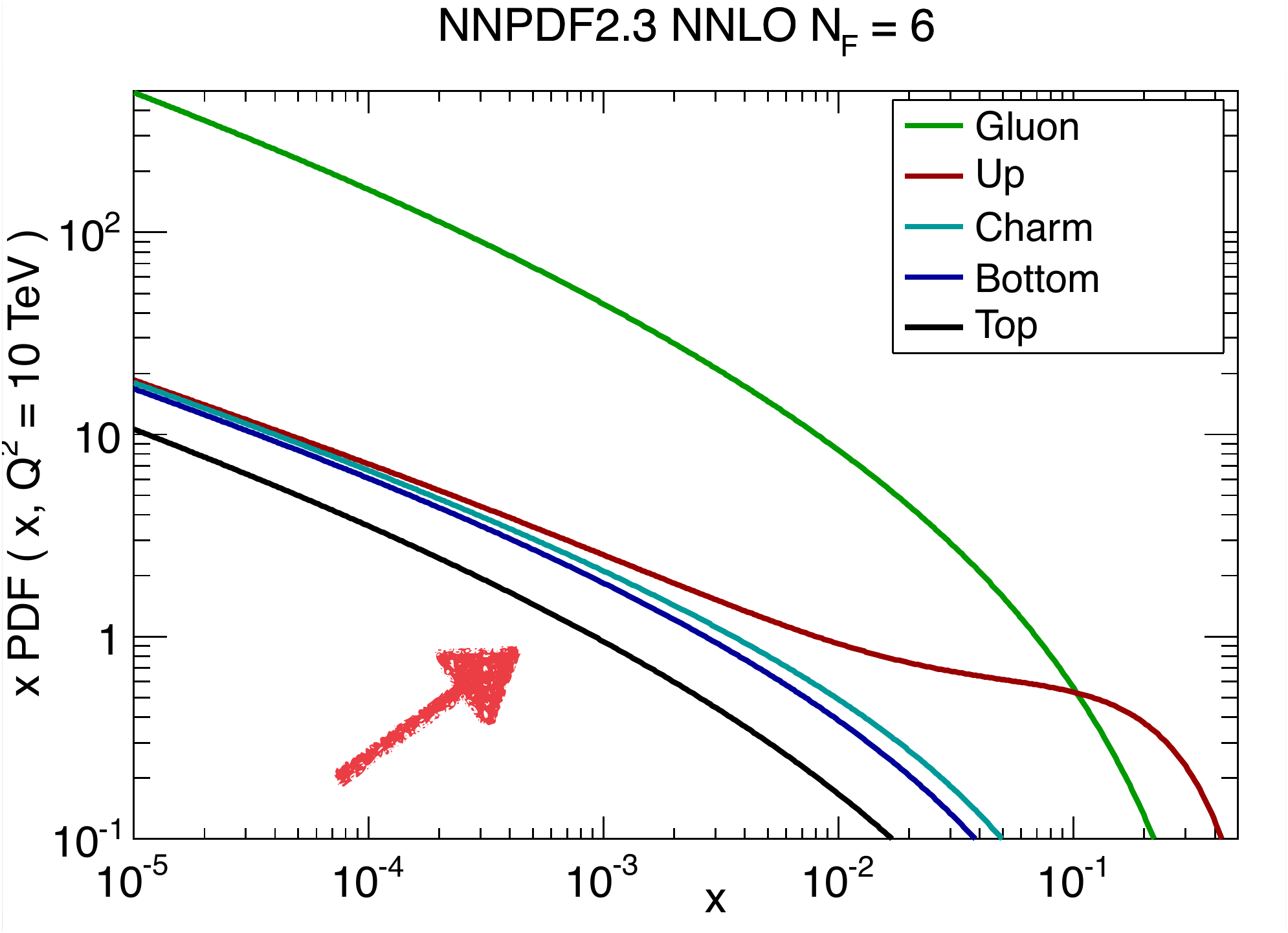}
  \includegraphics[width=0.51\textwidth,page=4]{figs/gluon_PDF_small-x_comparison_new.pdf}
  \caption{Left plot: the NNPDF2.3 NNLO $n_f=6$ PDF set at $Q=10$ TeV
    including the top quark PDF.
    Right plot: the gluon  and quark single PDFs
    obtained from a preliminary DIS-only fit with NLO+NLLx DGLAP evolution,
    compared to a baseline
    fit with NLO evolution.
  }
  \label{fig:gluonsmallxresRatio}
\end{figure}

\paragraph{Top as a massless parton.}
At a 100~TeV hadron collider,
for some processes 
the top quark mass can be much smaller
than the typical hard-scattering collision energy.
For $Q\sim 10$ TeV, for instance, $\alpha_s(Q)\log(Q^2/m_t^2)\sim 0.6$, which makes a
perturbative expansion of the hard process questionable.
Therefore, one might wonder if the concept of top quark PDF is relevant at the
FCC, just as charm and bottom PDFs are commonly used in LHC calculations.
As with charm and bottom,
introducing a PDF for the top quark 
allows to resum potentially large collinear logarithms of the form
$\alpha_s^n(Q)\log^n(Q^2/m_t^2)$ to all orders in perturbation
theory.
The generalization of the DGLAP evolution equations to include
a top PDF up to NNLO is straightforward, and indeed most modern PDF sets
provide variants where the maximum number of light quarks in the
PDF evolution is set to $n_f=6$.
In Fig.~\ref{fig:gluonsmallxresRatio} we show the top quark PDF,
evaluated at $Q=10$ TeV, together with the other
light partons, for the
NNPDF2.3NNLO $n_f=6$ PDF set~\cite{Ball:2012cx}.
We observe that the top quark
PDF can be of a similar size as the light
quark PDFs, in particular at medium and small-$x$.

So while technically
generating a top quark PDF is straightforward, it still needs to be
demonstrated if it provides any calculational advantage over using the standard
FFN scheme, where the top quark is treated as massive,
even for the extreme energies of a 100 TeV collider.
This issue has been recently studied in~\cite{Han:2014nja,Dawson:2014pea},
as well as the SM FCC report~\cite{report},
reaching similar conclusions:
a purely massless treatment of top quarks
is unreliable even at 100 TeV, but the concept of a top quark PDF
is still relevant in the context of matched calculations of heavy quark
production such as
FONLL~\cite{Cacciari:1998it}.

\paragraph{High-energy resummation at 100 TeV.}
\label{sec:pdf_resum}

When Bjorken-$x$ is small enough, logarithms of the form $\ln^k 1/x$
in the DGLAP splitting functions and in partonic matrix elements become
large, and might hamper the standard perturbative expansion.
These logarithms can be resummed to all orders by means
of the  $k_t$ factorization
theorem.
Ongoing work~\cite{BMPinpreparation}
aims at providing resummed anomalous dimensions and coefficient functions
through a fast \texttt{C++} code named \texttt{HELL} 
interfaced to  \texttt{APFEL}~\cite{Bertone:2013vaa}, able
 to perform DGLAP evolution with NLLx small-$x$ resummation
matched to the fixed order up to NLO.
Using this code, preliminary  DIS-only
NNPDF fits with NLLx evolution have been performed.
In Fig.~\ref{fig:gluonsmallxresRatio}
we show the ratio of the gluon and quark singlet PDFs
    obtained from a DIS-only fit with NLO+NLLx DGLAP evolution to the corresponding
    fit with NLO evolution at $Q=100$~GeV.
    In this comparison, NLO coefficient functions are used
    in both fits.
As compared to the fixed-order NLO fit, including NLLx small-$x$
effects in the PDF evolution leads to a suppression of the gluon and quark
singlet for $x\lsim 10^{-4}$, which reaches 20\% at $x\simeq 10^{-7}$,
and  moderate few-percent enhancement of the PDFs at intermediate
$x$.
The further suppression at large-$x$ is related to the lack of
hadronic data.
This preliminary results indicate that
small-$x$ resummation effects should
be relevant for precision physics at a 100 TeV collider, and thus
deserve further investigation in this context.

\paragraph{Acknowledgments.}
I am grateful to my collaborators in
the PDF section of the
report on {\it ``Physics at a 100 TeV pp collider: Standard Model processes''},
and in particular to Michelangelo Mangano and Giulia Zanderighi for many
comments and suggestions, as well as to Valerio
Bertone, Stefano Carrazza,
Davide Pagani, Gavin Salam, and Marco Zaro for providing
some of the plots in this contribution.\\
This work has been supported by an STFC Rutherford Fellowship
and Grant ST/K005227/1 and ST/M003787/1, and
by an European Research Council Starting Grant ``PDF4BSM''.

\providecommand{\href}[2]{#2}\begingroup\raggedright\endgroup


\providecommand{\href}[2]{#2}\begingroup\raggedright\begin{thebibliography}{100}

\bibitem{Rojo:2015acz}
J.~Rojo et~al., {\it {The PDF4LHC report on PDFs and LHC data: Results from Run
  I and preparation for Run II}},  {\em J. Phys.} {\bf G42} (2015) 103103,
  [\href{http://arxiv.org/abs/1507.00556}{{\tt arXiv:1507.00556}}].

\bibitem{Butterworth:2015oua}
J.~Butterworth et~al., {\it {PDF4LHC recommendations for LHC Run II}},  {\em J.
  Phys.} {\bf G43} (2016) 023001, [\href{http://arxiv.org/abs/1510.03865}{{\tt
      arXiv:1510.03865}}].

\bibitem{Accardi:2016ndt} 
  A.~Accardi {\it et al.},{\it Recommendations for PDF usage in LHC predictions,''}, \href{http://arxiv.org/abs/1603.08906 }{{\tt
      arXiv:1603.08906 }}.
  
\bibitem{Ball:2014uwa}
{\bf NNPDF} Collaboration, R.~D. Ball et~al., {\it {Parton distributions for
  the LHC Run II}},  {\em JHEP} {\bf 04} (2015) 040,
  [\href{http://arxiv.org/abs/1410.8849}{{\tt arXiv:1410.8849}}].

\bibitem{Dulat:2015mca}
S.~Dulat, T.-J. Hou, J.~Gao, M.~Guzzi, J.~Huston, P.~Nadolsky, J.~Pumplin,
  C.~Schmidt, D.~Stump, and C.~P. Yuan, {\it {New parton distribution functions
  from a global analysis of quantum chromodynamics}},  {\em Phys. Rev.} {\bf
  D93} (2016), no.~3 033006, [\href{http://arxiv.org/abs/1506.07443}{{\tt
  arXiv:1506.07443}}].

\bibitem{Harland-Lang:2014zoa}
L.~A. Harland-Lang, A.~D. Martin, P.~Motylinski, and R.~S. Thorne, {\it {Parton
  distributions in the LHC era: MMHT 2014 PDFs}},  {\em Eur. Phys. J.} {\bf
  C75} (2015), no.~5 204, [\href{http://arxiv.org/abs/1412.3989}{{\tt
      arXiv:1412.3989}}].

\bibitem{Alekhin:2013nda}
  S.~Alekhin, J.~Blumlein and S.~Moch, {\it
    The ABM parton distributions tuned to LHC data,},
  {\em Phys.\ Rev.\ D} {\bf 89}, no. 5, 054028 (2014)
  [\href{http://arxiv.org/abs/1310.3059}{{\tt
        arXiv:1310.3059}}].

\bibitem{report} M.~Mangano, G.~Zanderighi et~al.,
  {\it Physics at a 100 TeV pp collider: Standard Model processes},
  in preparation.

  \bibitem{Carrazza:2015hva}
S.~Carrazza, J.~I. Latorre, J.~Rojo, and G.~Watt, {\it {A compression algorithm
  for the combination of PDF sets}},  {\em Eur. Phys. J.} {\bf C75} (2015) 474,
  [\href{http://arxiv.org/abs/1504.06469}{{\tt arXiv:1504.06469}}].

\bibitem{Martin:2004dh}
A.~D. Martin, R.~G. Roberts, W.~J. Stirling, and R.~S. Thorne, {\it {Parton
  distributions incorporating QED contributions}},  {\em Eur. Phys. J.} {\bf
  C39} (2005) 155--161, [\href{http://arxiv.org/abs/hep-ph/0411040}{{\tt
  hep-ph/0411040}}].

\bibitem{Ball:2013hta}
{\bf NNPDF} Collaboration, R.~D. Ball, V.~Bertone, S.~Carrazza, L.~Del~Debbio,
  S.~Forte, A.~Guffanti, N.~P. Hartland, and J.~Rojo, {\it {Parton
  distributions with QED corrections}},  {\em Nucl. Phys.} {\bf B877} (2013)
  290--320, [\href{http://arxiv.org/abs/1308.0598}{{\tt arXiv:1308.0598}}].

\bibitem{Schmidt:2015zda}
C.~Schmidt, J.~Pumplin, D.~Stump, and C.~P. Yuan, {\it {CT14QED PDFs from
  Isolated Photon Production in Deep Inelastic Scattering}},
  \href{http://arxiv.org/abs/1509.02905}{{\tt arXiv:1509.02905}}.

\bibitem{Bertone:2013vaa}
V.~Bertone, S.~Carrazza, and J.~Rojo, {\it {APFEL: A PDF Evolution Library with
  QED corrections}},  {\em Comput. Phys. Commun.} {\bf 185} (2014) 1647--1668,
  [\href{http://arxiv.org/abs/1310.1394}{{\tt arXiv:1310.1394}}].

\bibitem{Alwall:2014hca}
J.~Alwall, R.~Frederix, S.~Frixione, V.~Hirschi, F.~Maltoni, et~al., {\it {The
  automated computation of tree-level and next-to-leading order differential
  cross sections, and their matching to parton shower simulations}},  {\em
  JHEP} {\bf 1407} (2014) 079, [\href{http://arxiv.org/abs/1405.0301}{{\tt
  arXiv:1405.0301}}].

\bibitem{Bertone:2015lqa}
V.~Bertone, S.~Carrazza, D.~Pagani, and M.~Zaro, {\it {On the Impact of Lepton
  PDFs}},  {\em JHEP} {\bf 11} (2015) 194,
  [\href{http://arxiv.org/abs/1508.07002}{{\tt arXiv:1508.07002}}].

\bibitem{Maltoni:2012pa}
F.~Maltoni, G.~Ridolfi, and M.~Ubiali, {\it {b-initiated processes at the LHC:
  a reappraisal}},  {\em JHEP} {\bf 07} (2012) 022,
  [\href{http://arxiv.org/abs/1203.6393}{{\tt arXiv:1203.6393}}]. [Erratum:
  JHEP04,095(2013)].

\bibitem{Ball:2012cx}
R.~D. Ball et~al., {\it {Parton distributions with LHC data}},  {\em Nucl.
  Phys.} {\bf B867} (2013) 244--289,
  [\href{http://arxiv.org/abs/1207.1303}{{\tt arXiv:1207.1303}}].

\bibitem{Han:2014nja}
T.~Han, J.~Sayre, and S.~Westhoff, {\it {Top-Quark Initiated Processes at
  High-Energy Hadron Colliders}},  {\em JHEP} {\bf 04} (2015) 145,
  [\href{http://arxiv.org/abs/1411.2588}{{\tt arXiv:1411.2588}}].

\bibitem{Dawson:2014pea}
S.~Dawson, A.~Ismail, and I.~Low, {\it {A Redux on "When is the Top Quark a
  Parton?"}},  {\em Phys.Rev.} {\bf D90} (2014), no.~1 014005,
  [\href{http://arxiv.org/abs/1405.6211}{{\tt arXiv:1405.6211}}].

\bibitem{Cacciari:1998it}
M.~Cacciari, M.~Greco, and P.~Nason, {\it {The P(T) spectrum in heavy flavor
  hadroproduction}},  {\em JHEP} {\bf 05} (1998) 007,
  [\href{http://arxiv.org/abs/hep-ph/9803400}{{\tt hep-ph/9803400}}].

\bibitem{BMPinpreparation}
M.~Bonvini, S.~Marzani, and T.~Peraro, {\it {in preparation}}.

\end{thebibliography}\endgroup
\end{document}